\documentclass[aps,prd,twocolumn,showpacs,superscriptaddress,nofootinbib,preprintnumbers]{revtex4-2}

\usepackage{style}

\begin{document}

%\preprint{FERMILAB-PUB-OMG-WTF-BBQ}

%\title{Stellar-Mass Black Holes with Dark Matter Mini-Spikes:\\Evidence of a Primordial Origin?}

\title{Dark Spiky Primordial Black Holes}

\author{Aurora Ireland\,\orcidlink{0000-0001-5393-0971}} 
\email{anireland@uchicago.edu}
\affiliation{Department of Physics, University of Chicago, Chicago, IL 60637}

\date{\today}

\begin{abstract}
Recent observations of two nearby black hole low-mass X-ray binaries have suggested possible indication of dark matter density spikes. While the evidence is compelling, one issue with this interpretation is that light black holes formed from stellar collapse are not expected to form dark matter spikes, and so it is unclear how the stellar-mass black holes in these binaries could have acquired such features. Since primordial black holes are expected to form ultra-dense dark matter mini-spikes, in this article we explore the possibility of a primordial origin for these light black holes.
\end{abstract}

\bigskip
\maketitle

\section{Introduction}\label{sec:intro}

Observations of two nearby black hole low-mass X-ray binaries (BH-LMXBs) have revealed anomalously fast orbital decay rates for the companion star. The BH-LMXBs A0620-00 and XTE J1118+480 each consist of an $\mathcal{O}(1\text{-}10)\, M_\odot$ black hole and a much lighter ($<M_\odot$) companion star with nearly circular orbit and short orbital period \cite{vanGrunsven:2017,Neilsen:2008,Hernandez:2013,Khargharia:2013,Wu:2015,Wu:2016,NovaMuscae:2017}. The dominant sources of angular momentum loss for such systems are magnetic braking \cite{Verbunnt:1981}, mass transfer from the donor star \cite{Rappaport:1982}, and gravitational wave emission, from which we expect a decay rate of $|\dot{P}| \lesssim 0.03 \, \text{ms}/\text{yr}$ \cite{NovaMuscae:2017}. This is significantly smaller than the measured\footnote{Based on observations made with the 10.4m Gran Telescopio Canarias equipped with the OSIRIS spectrograph \cite{OSIRIS}.} rates of $\dot{P} = -0.60 \pm 0.08 \, \text{ms}/\text{yr}$ and $\dot{P} = -1.90 \pm 0.57 \, \text{ms}/\text{yr}$, respectively \cite{Hernandez:2013,GonzalezHernandez:2011}. 

An extremely strong stellar magnetic field could in principle account for the decay of A0620-00, though this is inconsistent with the observed binary mass loss rate \cite{Hernandez:2013}. Other potential explanations have invoked angular momentum loss due to interactions between the circumbinary disk and the inner binary \cite{Xu:2018}. However the inferred mass transfer rate and circumbinary disk mass appear insufficient for this hypothesis to be viable \cite{Chen:2015}. At present, the origin of these rapid decays remains a mystery. 
%whose resolution likely requires the introduction of some as-yet-unknown element.  

The authors of Ref.~\cite{Chan:2022} recently put forth the possibility that dynamical friction due to a dark matter (DM) density spike around the BH may be responsible for the rapid orbital decays in these binaries. As the orbiting companion star traverses the spike, it pulls DM particles into its wake, resulting in a drag force which reduces its energy --- \textit{dynamical friction} \cite{Chandrasekhar:1943I,Chandrasekhar:1943II,Chandrasekhar:1943III}. Their analysis found density spikes with ``typical'' spike indices of $\gamma_{\rm sp} \sim 1.7-1.9$ to be consistent with the observed decay rates, leading them to suggest that these systems could provide indirect evidence of DM spikes. 

Their analysis is flawed, however, in that they assumed a DM profile appropriate for a massive BH residing in the galactic center, not a stellar-mass BH in a LMXB. DM spikes arise in the context of intermediate-mass and supermassive BHs growing adiabatically in cold, collisionless DM halos \cite{Gondolo:1999,Will:2013}. Due to the conservation of adiabatic invariants \cite{Quinlan:1995}, an initial cusp profile $\rho \propto r^{-\gamma}$ will be compressed to a new profile $\rho \propto r^{-\gamma_{\rm sp}}$ with greater spike index $\gamma_{\rm sp} = (9 - 2 \gamma)/(4 - \gamma)$ --- a \textit{dark matter spike}. 
%An initial cusp profile $\rho \propto r^{-\gamma}$ will evolve to a new profile $\rho \propto r^{-\gamma_{\rm sp}}$ with greater spike index $\gamma_{\rm sp} = (9 - 2 \gamma)/(4 - \gamma)$ due to the conservation of adiabatic invariants \cite{Quinlan:1995}. Such an adiabatically-compressed profile is called a \textit{dark matter spike}. 
Due to their much weaker gravitational influence, their locations in DM sparse regions, and the energetic nature of BH collapse, stellar-mass BHs of an astrophysical origin are not expected to form DM spikes.

Stellar-mass BHs of a primordial origin, however, \textit{are} expected to efficiently accrete DM particles, forming high-density mini-spikes \cite{Bertschwinger:1985,Mack:2006,Ricotti:2007}. This begins soon after primordial black hole (PBH) formation, as DM particles which have already decoupled from the Hubble flow fall into the PBH's gravitational sphere of influence \cite{Eroshenko:2016}. Following matter-radiation equality, the spike continues to grow through the accretion and virialization of DM \cite{Bertschwinger:1985}. The resultant DM profile interpolates between a compact constant density core and a power law $\rho \propto r^{-9/4}$ \cite{Boucenna:2017,Adamek:2019,Carr:2020}. This profile may be modified by DM annihilations as well as astrophysical processes such as tidal stripping, dynamical friction, gravitational scattering, and more. 

In this article, we investigate whether a primordial origin for the BHs in these LMXBs can account for the DM density inferred by the dynamical friction interpretation of the anomalous orbital decays. 
%We begin in Sec.~\ref{sec:densityprofile} by reviewing DM mini-spike formation and growth before comparing these profiles with the DM density implied by the dynamical friction hypothesis in Sec.~\ref{sec:confrontingthedata}.

%%%%%%%%%%%%%%%%%%%%%%%%%%%%%%%%%%%%%%%%%%%%%%%%%%%%%%%%%%%%%%
%%%%%%%%%%%%%%%%%%%%%%%%%%%%%%%%%%%%%%%%%%%%%%%%%%%%%%%%%%%%%%
\section{Halo Density Profile}\label{sec:densityprofile}

%%%%%%%%%%%%%%%%%%%%%%%%%%%%%%%%%%%%%%%%%%%%%%%%%%%%%%%%%%%%%%
\subsection{Formation during Radiation Domination}\label{subsec:raddensityprofile}

Mini-spike assembly begins during radiation domination, soon after PBH formation \cite{Eroshenko:2016}. Cold DM particles with low velocities in the Boltzmann tail will decouple from the Hubble flow, falling into finite orbits about the PBH. Since the DM mass accreted during radiation domination is at most some $\mathcal{O}(1)$ fraction of the PBH mass, the gravitational potential of the PBH will dominate throughout this era. We thus approximate the Newtonian\footnote{Since general relativistic effects are important only out to a radius $r \sim 10 r_s$ \cite{Will:2013} and we are interested in large radii $r \gg r_s$ near the orbiting body, we adopt a Newtonian analysis and remain agnostic as to the precise DM profile near the horizon.} potential as $\Phi(r) = - G M_{\rm BH}/ r$. 

Presuming a FLRW background with scale factor $a$, the radial distance between a given DM particle and the PBH evolves as
\begin{equation}
    \ddot{r} = \frac{\ddot{a}}{a} r - \frac{G M_{\rm BH}}{r^2} \,.
\end{equation}
The particle decouples from the background expansion once the second term exceeds the first. We can use this condition to derive an approximate expression for the \textit{turn-around time} $t_{\rm ta}$ --- the time at which $\dot{r} = 0$ and the particle begins moving towards the PBH. This is related to the turn-around radius as $r_{\rm ta} \simeq (2 r_s)^{1/3} t_{\rm ta}^{2/3}$, where $r_s = 2 G M_{\rm BH}$ is the Schwarzschild radius. The numerical simulations of Ref.~\cite{Adamek:2019} suggest that a more accurate estimate is
\begin{equation}\label{eq:rta}
    r_{\rm ta} \simeq 1.0 \, r_s^{1/3} \, t_{\rm ta}^{2/3} \,,
\end{equation}
which we adopt going forward. This radius can also be understood as defining the PBH's sphere of influence.

%%%%%%%%%%%%%%%%%%%%%%%%%%%%%%%%%%%%%%%%%%%%%%%%%%%%%%%%%%%%%%
\subsubsection{The Initial Profile}

The initial DM profile depends on whether or not the DM has kinetically decoupled at the time of PBH formation \cite{Adamek:2019,Carr:2020}. The initial mass of a PBH formed by the collapse of a primordial overdensity upon horizon re-entry is set by the mass of the comoving horizon $M_H = (2 G H)^{-1}$ as $M_{\rm BH} = \gamma_{\rm eff} M_H$, where $\gamma_{\rm eff} \simeq 0.2$ during radiation domination. The temperature at formation\footnote{Note that $\mathcal{O}(1-10) \, M_\odot$ PBHs form around the QCD phase transition, a particularly attractive window for PBH formation due to the softening of the equation of state \cite{Musco:2023}.} is then
\begin{equation}
    T_{\rm form} = 141\, \text{MeV} \left( \frac{\gamma_{\rm eff}}{0.2} \right)^{1/2} \left( \frac{24.0}{g_\star(T)} \right)^{1/4} \left( \frac{M_\odot}{M_{\rm BH}} \right)^{1/2} \,,
\end{equation}
where $g_\star(T)$ is the effective degrees of freedom. This should be compared with the temperature $T_{\rm kd}$ at which interactions fail to keep the dark matter in kinetic equilibrium with the Standard Model. Note that unlike Refs.~\cite{Adamek:2019,Carr:2020}, we do not restrict to WIMPs, and thus treat $T_{\rm kd}$ as a free parameter. 

For PBHs which form prior to kinetic decoupling, the DM is initially too tightly coupled to the primordial plasma to appreciably accrete. Consequently, the PBH is simply surrounded by a DM halo with constant background density $\rho_{\rm kd}$ extending out to $r_{\rm ta}(t_{\rm kd})$, where
\begin{equation}\label{eq:rhokd}
    \rho_{\rm kd} = \frac{\rho_{\rm eq}}{2} f_{\rm DM} \left( \frac{t_{\rm eq}}{t_{\rm kd}} \right)^{3/2} \,,
\end{equation}
and $f_{\rm DM}$ the \textit{particle} DM fraction\footnote{In this scenario, the DM is composed of both particle DM and PBHs, $\rho_{\rm DM,tot} = \rho_{\rm DM} + \rho_{\rm PBH}$, with fractional contributions $f_{\rm DM} \equiv \rho_{\rm DM}/\rho_{\rm DM,tot}$ and $f_{\rm PBH} \equiv \rho_{\rm PBH}/\rho_{\rm DM,tot}$.}. Following kinetic decoupling, DM can be more effectively captured by the PBH and the halo radius grows. Since the density at $r_{\rm ta}$ is close to the background DM density, which dilutes as $\rho \propto a^{-3} \propto t^{-3/2}$, the density of the halo beyond $r_{\rm ta}(t_{\rm kd})$ falls. Using the scaling in Eq.~(\ref{eq:rta}), the profile for radii $r_{\rm ta}(t_{\rm kd}) \leq r \leq r_{\rm ta}(t_{\rm eq})$ is the \textit{mini-spike} profile
\begin{equation}\label{eq:rhospike}
    \rho_{\rm sp}(r) = \frac{\rho_{\rm eq}}{2} f_{\rm DM} \left( \frac{r_{\rm ta}(t_{\rm eq})}{r} \right)^{9/4} \,.
\end{equation}
The radial profile thus interpolates between a dense, constant core\footnote{Note that this core does not extend all the way to the Schwarzschild radius, but rather must vanish at $r=2 r_s$ \cite{Will:2013}.} at small $r$ and spike profile at larger $r$,
\begin{equation}\label{eq:tilderhoi}
    \rho_i(r) = \begin{cases} \rho_{\rm kd} & r_s < r \leq r_{\rm ta}(t_{\rm kd}) \,, \\ \rho_{\rm sp}(r) & r_{\rm ta}(t_{\rm kd}) \leq r \leq r_{\rm ta}(t_{\rm eq}) \,. \end{cases}
\end{equation}
In the second case of a PBH forming \textit{after} kinetic decoupling, there is no longer a constant density core, and the profile is simply $\rho_i(r) = \rho_{\rm sp}(r)$.

\begin{comment}
Note that this $\rho_{\rm sp} \propto r^{-9/4}$ scaling can also be understood as a consequence of adiabatic compression of the DM halo. For a BH growing adiabatically in a pre-existing DM halo with $\rho \propto r^{-\gamma}$, the conservation of radial action and angular momentum imply a final density profile which is also a power law $\rho \propto r^{-\gamma_{\rm sp}}$, with \cite{Quinlan:1995,Gondolo:1999,Will:2013}
\begin{equation}
    \gamma_{\rm sp} = \frac{9 - 2 \gamma}{4 - \gamma} \,.
\end{equation}
In our case, the initial DM density about the PBH is uniform ($\gamma = 0$), leading to $\gamma_{\rm sp} = 9/4$.
\end{comment}

%%%%%%%%%%%%%%%%%%%%%%%%%%%%%%%%%%%%%%%%%%%%%%%%%%%%%%%%%%%%%%
\subsubsection{Modification by Orbital Motion of DM}

\begin{comment}
A large thermal kinetic energy $K$ can prevent the formation of a bound halo, so we should ensure that it is small relative to the potential energy $U$. The ratio $|K/U|$ is greatest at turn-around, so it suffices to check that $|K(r_{\rm ta})/U(r_{\rm ta})| < 1$ \cite{Adamek:2019}. 
%The kinetic energy is set by the DM temperature $K \sim T_{\rm DM}$, which falls as $\propto a^{-2}$ after kinetic decoupling, while the potential energy is $U = - G m_{\rm DM} M_{\rm BH}/r$. 
This corresponds to the condition $x_{\rm ta} \equiv r_{\rm ta}/r_s > x_K$, defining
\begin{equation}
\begin{split}
    x_{K} & = (2.22 \times 10^{-4}) \\ & \times \left( \frac{10.8}{g_\star(T_{\rm kd})} \right) \left( \frac{10\, \text{MeV}}{T_{\rm kd}} \right)^2 \left( \frac{M_\odot}{M_{\rm BH}} \right)^2 \left( \frac{1\, \text{TeV}}{m_{\rm DM}} \right)^2 \,.
    \end{split}
\end{equation}
This inequality can be satisfied for our $\mathcal{O}(1\text{-}10)\, M_\odot$ PBHs provided the DM is sufficiently massive, in which case we are justified in neglecting the potentially disruptive effect of the DM's thermal kinetic energy. 
\end{comment}

The finite velocities and orbital motions of the DM particles will modify the density profile. This effect was accounted for in Ref.~\cite{Eroshenko:2016} by using conservation of phase space to write
\begin{equation}\label{eq:rhonum}
    \rho(r) = \frac{2}{r^2} \int d^3 v_i \, f_B(v_i) \int dr_i \, r_i^2 \, \frac{\rho_i(r_i)}{\tau_{\rm orb}} \left| \frac{dt}{dr} \right| \,,
\end{equation}
where $r_i$ is the initial radial distance at turn-around, $v_i$ is the initial velocity, $f_B(v_i)$ is the DM velocity distribution, $\tau_{\rm orb}$ is the DM orbital period, and $\rho_i$ is the initial density profile of Eq.~(\ref{eq:tilderhoi}). See Appendix~\ref{sec:appendixA} for more detail. In the limit of cold DM with small velocities, Eq.~(\ref{eq:rhonum}) reduces to
\begin{equation}
    \rho(x) \simeq \frac{2}{\pi} \int_x^\infty dx_i \frac{x_i}{x^{3/2}} \frac{\rho_i(x_i)}{\sqrt{x_i - x}} \,,
\end{equation}
where we have introduced the dimensionless $x \equiv r/r_s$. In the region $x> x_{\rm ta}(t_{\rm kd}) \equiv x_{\rm kd}$, this evaluates to 
\begin{equation}\label{eq:rhoamp}
    \rho(x) \simeq \alpha \, \rho_{\rm kd} \left( \frac{x_{\rm kd}}{x} \right)^{9/4} \,,
\end{equation}
with $\alpha \equiv \sqrt{\frac{4}{\pi}} \frac{\Gamma(3/4)}{\Gamma(5/4)} \simeq 1.526$. Thus, the orbital motion of DM simply serves to amplify the density everywhere by $\sim 53\%$, with the power law scaling $\rho \propto x^{-9/4}$ intact.

%%%%%%%%%%%%%%%%%%%%%%%%%%%%%%%%%%%%%%%%%%%%%%%%%%%%%%%%%%%%%%
\subsubsection{Modification by DM Annihilations}

For self-annihilating DM, the rate scales as $\Gamma_{\rm ann} \propto \rho^2$ and so is strongly enhanced in the dense inner core of the mini-spike, leading to an amplified $\gamma$-ray emission signal \cite{Gondolo:1999}. Bounds on the allowed galactic and extragalactic $\gamma$-ray flux then allow us to constrain this hybrid scenario of both PBH and particle DM \cite{Boucenna:2017,Carr:2020}. In particular for WIMP DM with weak-scale annihilation cross sections, this scenario has been demonstrated to be incompatible with stellar-mass PBHs \cite{Adamek:2019}.

Self-annihilations modify the density profile since they impose an upper limit on the DM density \cite{Ullio:2002,Scott:2009,Carr:2020}
\begin{comment}
\begin{equation}
    \rho_{\rm max} \simeq \frac{f_{\rm DM} m_{\rm DM} H}{\expval{\sigma v}} \,,
\end{equation}
\end{comment}
\begin{equation}
    \rho_{\rm max} \simeq (1.3 \times 10^{-16}\, \text{g}/\text{cm}^3) f_{\rm DM} \left( \frac{m_{\rm DM}}{\text{GeV}} \right) \left( \frac{3 \times 10^{-26}\, \text{cm}^3/\text{s}}{\expval{\sigma v}} \right) 
\end{equation}
where $\expval{\sigma v}$ is the velocity-weighted annihilation cross section. Equating this with Eq.~(\ref{eq:rhoamp}) allows us to define the radius $x_{\rm ann}$ below which the power law gives way to a constant density plateau
\begin{equation}
    x_{\rm ann}(t) = x_{\rm eq} \left( \frac{\alpha}{2} \frac{\rho_{\rm eq}}{m_{\rm DM}} \frac{\expval{\sigma v}}{H(t)} \right)^{4/9} \,,
\end{equation}
where $x_{\rm eq} \equiv x_{\rm ta}(t_{\rm eq})$. The modified profile is thus
\begin{equation}\label{eq:annmodifiedrho}
    \rho_{\rm ann}(x) = \begin{cases} \rho_{\rm max} & 2 < x \leq x_{\rm ann} \,, \\ \rho_{\rm max} \left(\frac{x_{\rm ann}}{x} \right)^{9/4} & x_{\rm ann} \leq x \leq x_{\rm eq} \,. \end{cases} 
\end{equation}
Closer examination has since revealed that instead of a flat annihilation plateau, the spike is flattened to a weak cusp with either $\rho \propto r^{-1/2}$ for s-wave annihilations \cite{Vasiliev:2007} or $\rho \propto r^{-0.34}$ for p-wave \cite{Shapiro:2016}. 

We do not consider annihilating DM further, as the parameter space needed to explain the anomalous orbital decays is already ruled out on the basis of $\gamma$-ray overproduction. See Appendix~\ref{sec:appendixB} for more details.

%%%%%%%%%%%%%%%%%%%%%%%%%%%%%%%%%%%%%%%%%%%%%%%%%%%%%%%%%%%%%%
\subsection{Growth during Matter Domination}\label{subsec:matterdensityprofile}

Following matter-radiation equality, accretion becomes efficient and the mass of the DM halo quickly grows to exceed that of the PBH. The dominant process contributing to growth during this era is \textit{secondary infall}, wherein bound shells of infalling DM are added at successively larger radii \cite{Bertschwinger:1985}. The infall process assumes a self-similar form and for cold, collisionless DM results in a power law profile $\rho \propto r^{-9/4}$ \cite{Bertschwinger:1985,Mack:2006,Adamek:2019} --- the same radial dependence as the initial mini-spike formed during radiation domination.

To build intuition for this scaling, note that the PBH-halo system constitutes an initial overdensity $\Delta = \delta M/M_{\rm bg}$, where $\delta M$ is the combined mass of the PBH-halo system and $M_{\rm bg} = \frac{4\pi}{3} \rho_{\rm bg} R^3$ is the background mass in the region \cite{Carr:2020}. Since density perturbations grow as $\Delta \propto a \propto t^{2/3}$ during matter domination, the mass gravitationally bound to the PBH-halo system grows as
\begin{equation}\label{eq:boundmass}
    M_{\rm bound}(t) = M_{\rm bound}(t_{\rm eq}) \left( \frac{t}{t_{\rm eq}} \right)^{2/3} \,.
\end{equation}
This can amount to a factor $\sim 100$  growth between matter-radiation equality and the onset of non-linear structure formation \cite{Mack:2006}. Meanwhile using that the total energy density dilutes as $\rho \propto a^{-3} \propto t^{-2}$, the turn-around radius of a DM shell binding at time $t$ is
\begin{equation}
    r \propto \left( \frac{M_{\rm bound}(t)}{\rho(t)} \right)^{1/3} \propto t^{8/9} \propto \rho^{-4/9} \,,
\end{equation}
which can be inverted to obtain $\rho \propto r^{-9/4}$. Thus we recover the same power law scaling as before, and the density profile of Eq.~(\ref{eq:rhospike}) should be extended beyond $r_{\rm ta}(t_{\rm eq})$
\begin{equation}\label{eq:rhomatt}
    \rho(x) \simeq \begin{cases} \alpha \, \rho_{\rm kd} & 1 < x \leq x_{\rm kd} \,, \\ \alpha \, \rho_{\rm sp}(x) & x \geq x_{\rm kd} \,. \end{cases}
\end{equation}

%%%%%%%%%%%%%%%%%%%%%%%%%%%%%%%%%%%%%%%%%%%%%%%%%%%%%%%%%%%%%%
\subsection{Modification by Astrophysical Effects}\label{subsec:modification}

Various astrophysical processes can modify or disrupt the DM mini-spike, particularly at late times. These include tidal stripping \cite{Hertzberg:2019,Schneider:2010}, gravitational scattering and other interactions with stars in dense baryonic environments \cite{Ullio:2001,Bertone:2005,Shapiro:2022}, dynamical friction \cite{Kavanagh:2020,Coogan:2021}, mergers with other BHs \cite{Nishikawa:2017,Jangra:2023}, and more. Of course, the BH-LMXB formation event itself may also disrupt the mini-spike. 

Though most evolve from stellar binaries, BH-LMXBs may also form through dynamical capture, wherein a star passing by a BH loses energy through gravitational interactions and becomes gravitationally bound. The likelihood of dynamical capture is highest in dense stellar environments, like globular clusters or galactic centers, as well as when the BH is significantly more massive than the star, as in our BH-LMXBs of interest. 

For PBHs, there is also energy dissipation due to dynamical friction as the star passes through the DM mini-spike, which greatly facilitates capture. In Appendix~\ref{sec:appendixB2}, we estimate the number of binaries expected to form in this way, finding the number to be surprisingly large for our parameter space of interest. For example, presuming an abundance $f_{\rm PBH} \sim 10^{-5}$ of $\sim 5 \, M_\odot$ PBHs, one would expect $\sim 10^6$ binaries in the Milky Way. 

Interestingly, there is strong evidence that XTE J1118+480 formed through dynamical capture. Its stellar component has a metal-rich composition consistent with pollution from a supernova event, which would have ejected the companion from its system of origin \cite{GonzalezHernandez:2006,Hernandez:2008}. The members of this binary were then likely not born together, instead first encountering one another following the kick and binding dynamically. 

Because the binding energy of the spike always far exceeds the energy lost to dynamical friction (see Appendix~\ref{sec:appendixB2}), we are assured that the mini-spike is not destroyed during binary formation. Since formation often occurs in dense stellar environments, we should also consider gravitational scattering, tidal stripping, and other interactions with stars. Ref.~\cite{Delos:2022} demonstrated that the highly compact nature of PBH mini-spikes makes them particularly resistant to these sorts of disruptive effects. 

We anticipate that dynamical friction and gravitational scattering with the companion star will be the dominant effect modifying the DM profile. Since most of the cold DM particles have speeds less than the star's orbital speed $\nu$, they will gain energy through these processes. The energy lost by the star will go into heating up the DM halo, increasing the velocity dispersion and leading to a localized decrease in DM density.

This intuition is confirmed by the numerical simulations of Ref.~\cite{Kavanagh:2020}, which demonstrate in the context of BH-BH binaries that dynamical friction can reduce the DM density at the orbital radius by orders of magnitude. Using a modified version of the \texttt{HaloFeedback} code of \cite{HaloFeedback:2020} (see Appendix~\ref{sec:appendixC}), we include in Fig.~\ref{fig:mainresult} density profiles accounting for dynamical friction. We see that the DM density at the orbital radius is depleted by a factor $\sim 10^4$ through these scatterings.

%%%%%%%%%%%%%%%%%%%%%%%%%%%%%%%%%%%%%%%%%%%%%%%%%%%%%%%%%%%%%%
%%%%%%%%%%%%%%%%%%%%%%%%%%%%%%%%%%%%%%%%%%%%%%%%%%%%%%%%%%%%%%
\section{Confronting the Data}\label{sec:confrontingthedata}

The companion star traversing the DM spike will lose energy to dynamical friction, leading to orbital decay. In Ref.~\cite{Chan:2022}, it was argued that this effect could account for the anomalously fast orbital decays of A0620-00 and XTE J1118+480. We review here how the DM density can be inferred from the orbital decay rate and other precisely measured parameters organized in Table~\ref{tab:LMXBparams}.

\begin{table}[t!]
\centering
\begin{tabular}{|c||c|c|}
    \hline &&\\[-1em]
     & \textbf{A0620-00} & \,\,\textbf{XTE J1118+480} \\ 
    \hline\hline &&\\[-1em]
    $M_{\rm BH}$ ($M_\odot$) & $5.86 \pm 0.24$ \cite{vanGrunsven:2017} & $7.46^{+0.34}_{-0.69}$ \cite{Hernandez:2013} \\
    \hline &&\\[-1em]
    \,\,\,$q = M_*/M_{\rm BH}$\,\,\, & $0.060 \pm 0.004$ \cite{vanGrunsven:2017} & $0.024 \pm 0.009$ \cite{Khargharia:2013} \\
    \hline &&\\[-1em]
    $K$ (km/s) & $435.4 \pm 0.5$ \cite{Neilsen:2008} & $708.8 \pm 1.4$ \cite{Khargharia:2013} \\
    \hline &&\\[-1em]
    $i$ & $(54.1 \pm 1.1)^\circ$ \cite{vanGrunsven:2017} & $(73.5 \pm 5.5)^\circ$ \cite{Khargharia:2013} \\
    \hline &&\\[-1em]
    $P$ (day) & $0.32301415(7)$ \cite{Hernandez:2013} & $0.16993404(5)$ \cite{Hernandez:2013} \\
    \hline &&\\[-1em]
    $\dot{P}$ (ms/yr) & $-0.60 \pm 0.08$ \cite{Hernandez:2013} & $-1.90 \pm 0.57$ \cite{Hernandez:2013} \\
    \hline\hline &&\\[-1em]
    $x_{\rm orb}$ & $(1.46 \pm 0.04) \times 10^5$ & $8.01^{+0.56}_{-0.26} \times 10^{4}$ \\
    \hline &&\\[-0.9em]
    \,\, $\rho_{\rm orb} (\text{g}/\text{cm}^3)$ \,\, & $7.62^{+1.62}_{-1.42} \times 10^{-13}$ & $1.59^{+1.51}_{-0.74} \times 10^{-11}$ \\
    \hline 
\end{tabular}
\caption{Measured (top) and derived (bottom) parameters of the BH-LMXBs A0620-00 and XTE J1118+480.}
\label{tab:LMXBparams}
\end{table}

The star's energy loss due to dynamical friction proceeds at a rate \cite{Chandrasekhar:1943I,Dosopoulou:2023,Chan:2022}
\begin{equation}\label{eq:Elossrate}
    \dot{E}_* = - 4 \pi \mu^2 G^2 \rho_{\rm orb} \frac{\xi(\nu)}{\nu} \ln \Lambda \,,
\end{equation}
where $\mu = M_{\rm BH} M_*/(M_{\rm BH} + M_*)$ is the reduced mass, $\rho_{\rm orb} \equiv \rho(r_{\rm orb})$ is the DM density at the orbital radius, $\nu$ is the orbital speed, $\xi(\nu)$ characterizes the fraction of DM particles with speeds less than $\nu$, and $\ln \Lambda$ is the Coulomb logarithm. The measured parameters of Table~\ref{tab:LMXBparams} allow us to obtain $\nu$ from the radial velocity $K$ and orbital inclination $i$ as $\nu = K/\sin(i)$. We follow Ref.~\cite{Kavanagh:2020} in approximating $\Lambda \simeq \sqrt{M_{\rm BH}/M_*}$ and $\xi(\nu) \sim 1$. From the total mechanical energy of the system 
\begin{equation}\label{eq:totalE}
    E_* = - \frac{G M_{\rm BH} M_*}{2 r_{\rm orb}} \,,
\end{equation}
and Kepler's law for the period $P$
\begin{equation}
    P^2 = \frac{4\pi^2 r_{\rm orb}^3}{G(M_{\rm BH} + M_*)} \,,
\end{equation}
we obtain the following expression for the DM density at $r_{\rm orb}$ 
\begin{equation}\label{eq:rhomeasured}
    \rho_{\rm orb} = \frac{1}{6\pi} \frac{M_{\rm Pl}^{8/3}}{M_{\rm BH}^{1/3}} \frac{(1+q)^{5/3}}{q \ln (1/q)} \frac{\nu}{\xi(\nu)} \bigg| \frac{\dot{P}}{P} \bigg| \left( \frac{2\pi}{P} \right)^{2/3} \,.
\end{equation}  
From Table~\ref{tab:LMXBparams}, we infer $\rho_{\rm orb} \simeq 7.62^{+1.62}_{-1.42} \times 10^{-13}\, \text{g}/\text{cm}^3$ for A0620-00 and $\rho_{\rm orb} \simeq 1.59^{+1.51}_{-0.73} \times 10^{-11}\, \text{g}/\text{cm}^3$ for XTE J1118+480. 

\begin{figure}[t!]
\centering
\includegraphics[width=0.44\textwidth]{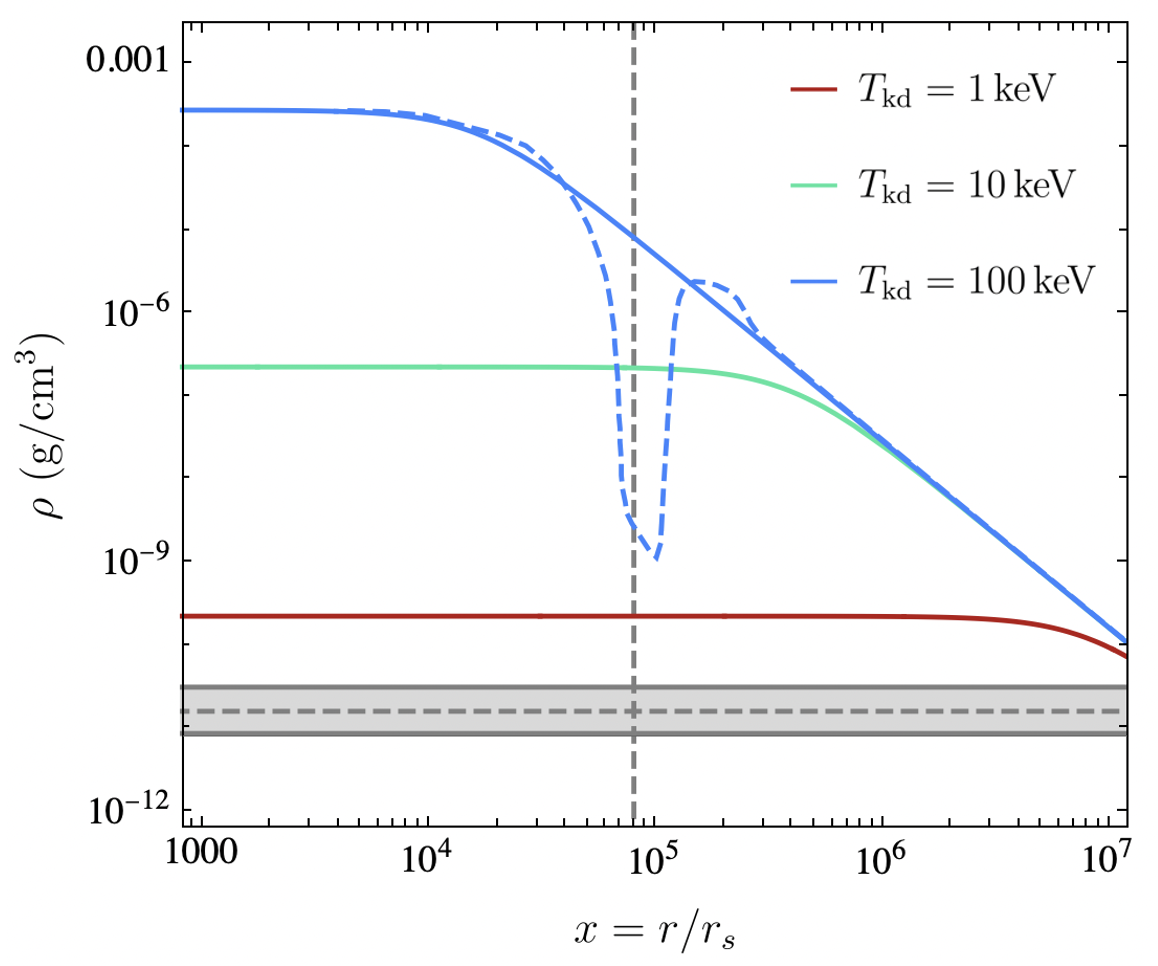}
\hspace{5mm}
\caption{Sample DM mini-spike profiles for a $7.46\, M_\odot$ PBH for DM kinetic decoupling temperatures of $1\,\text{keV}$ (red), $10\,\text{keV}$ (green), and $100\, \text{keV}$ (blue). Dashed lines indicate profiles altered by dynamical friction. The grey vertical dashed line denotes the orbital radius of XTE J1118+480 and the grey shaded horizontal region is the DM density inferred from dynamical friction.}
\label{fig:mainresult}
\end{figure}

Fig.~\ref{fig:mainresult} shows sample profiles for the DM mini-spike around a $7.46\, M_\odot$ PBH. 
%Downward arrows indicate that because we have neglected various astrophysical effects, these lines should be interpreted as upper bounds on $\rho$. 
We see that prior to taking into account halo feedback from dynamical friction, a PBH mini-spike could account for the inferred DM density so long as the DM has a relatively low\footnote{See Ref.~\cite{Bringmann:2016} for model-building efforts with late-decoupling DM.} kinetic decoupling temperature of $T_{\rm kd} \lesssim 1\, \text{keV}$. Dashed lines indicate profiles modified by dynamical friction backreaction. The observed density depletion at the orbital radius is consistent with the intuition of Sec.~\ref{subsec:modification}, and would make it such that a DM candidate with an earlier kinetic decoupling could account for the inferred density. The plot for A0620-00 is completely analogous.

\section{Conclusion}

If one interprets the anomalously fast orbital decays in these BH-LMXBs as originating from dynamical friction due to a DM spike, then one is led to the need for some mechanism to build up the DM spike, as stellar-mass BHs of an astrophysical origin are not expected to have them. Since PBHs \textit{are} expected to form high-density mini-spikes, in this article we have explored the possibility of a primordial origin for the BHs residing in these LMXBs. We have computed the mini-spike profile for a cold, collisionless DM candidate and found that for late kinematic decoupling times, this scenario could indeed account for the density inferred by the dynamical friction hypothesis. 

We leave to future work the numerical simulations needed to better model mini-spike disruption due to astrophysical effects. Beyond the mini-spike's impact on orbital dynamics, there are a few other potential signatures of this scenario. DM could affect the density and temperature profiles of the accretion disk, altering properties of the thermal emission and X-ray spectra. The mini-spike would also modify gravitational waves emitted by the binary, both through changes to the gravitational potential and dephasing due to dynamical friction \cite{Eda:2014,Kavanagh:2020,Coogan:2021}. Finally, the mini-spike may alter the BH's optical appearance, in particular the shadow and photon ring \cite{Macedo:2024,Faraji:2024}. Observations across multiple wavelengths would greatly aid in giving a more complete picture.

\section*{Acknowledgements}

AI thanks Robert Wagoner for first suggesting dark matter spikes as a possible topic of interest, for comments on an earlier version of this manuscript, and for many productive conversations. AI also thanks the organizers and participants of the New Horizons in Primordial Black Hole Physics workshop for constructive feedback. Conversations with Daniele Gaggero and Nicolas Esser were particularly helpful.

\appendix
%%%%%%%%%%%%%%%%%%%%%%%%%%%%%%%%%%%%%%%%%
%%%%%%%%%%%%%%%%%%%%%%%%%%%%%%%%%%%%%%%%%
\section{Modification by DM Orbital Motion}\label{sec:appendixA}

Here we determine how the initial density profile,
\begin{equation}\label{eq:tilderhoi}
    \rho_i(r) = \begin{cases} \rho_{\rm kd} & r_s < r \leq r_{\rm ta}(t_{\rm kd}) \,, \\ \rho_{\rm sp}(r) & r_{\rm ta}(t_{\rm kd}) \leq r \leq r_{\rm ta}(t_{\rm eq}) \,, \end{cases}
\end{equation}
is modified by the finite velocities and orbital motion of the DM. Following the treatment of Refs.~\cite{Eroshenko:2016,Carr:2020}, we begin by using conservation of phase space to write
\begin{equation}\label{eq:rhoeq}
    \rho(r) = \frac{2}{r^2} \int d^3 v_i \, f_B(v_i) \int dr_i \, r_i^2 \, \frac{\tilde{\rho}_i(r_i)}{\tau_{\rm orb}} \left| \frac{dt}{dr} \right| \,,
\end{equation}
where $r_i$ is the initial radial distance from the PBH at turn-around, $v_i$ is the initial velocity, $\tau_{\rm orb}$ is the DM orbital period, and $\rho_i$ is the initial density profile. We take the DM velocity to have a Maxwell-Boltzmann distribution, 
\begin{equation}\label{eq:boltzmann}
    f_B(v_i) = \left( \frac{m_{\rm DM}}{2 \pi T_{\rm DM}} \right)^{3/2} \exp \left( - \frac{m_{\rm DM} v_i^2}{2 T_{\rm DM}} \right) \,,
\end{equation}
where $T_{\rm DM}$ is the DM temperature, which depends on radial position as
\begin{equation}
    T_{\rm DM}(r) = T_{\rm kd} \left( \frac{r_{\rm ta}(t_{\rm kd})}{r} \right)^{3/2} \,.
\end{equation}
For convenience, we normalize radii to $r_s$ and define the dimensionless $x \equiv r/r_s$, $x_i \equiv r_i/r_s$. The initial energy of a DM particle then reads
\begin{equation}
    E_i = \frac{m_{\rm DM}}{2} \left( v_i^2 - \frac{1}{x_i} \right) \,,
\end{equation}
while the energy at an arbitrary time is
\begin{equation}
    E = \frac{m_{\rm DM}}{2} \left( \dot{r}^2 + \left(\frac{x_i}{x} \right)^2 v_i^2 \sin^2 \theta_i - \frac{1}{x} \right) \,,
\end{equation}
which we have simplified using the conservation of angular momentum $L = L_i = m_{\rm DM} r_i v_i \sin \theta_i$. Conservation of energy then allows us to identify the radial speed as
\begin{equation}\label{eq:rdot}
    \frac{dr}{dt} = \sqrt{v_i^2 - \frac{1}{x_i} + \frac{1}{x} - \left( \frac{x_i}{x} \right)^2 v_i^2 \sin^2 \theta_i} \,,
\end{equation}
and the orbital period as
\begin{equation}
    \tau_{\rm orb} = \pi r_s \left( \frac{x_i}{1 - x_i v_i^2} \right)^{3/2} \,.
\end{equation}
Writing $d^3 v_i = 2\pi v_i^2 dv_i d(\cos \theta_i)$ and performing the integral over $\cos \theta_i$, we arrive at the expression
\begin{equation}\label{eq:rhoinprogress}
    \rho(x) = \frac{4}{x} \int dx_i \, \frac{\rho_i(x_i)}{\sqrt{x_i}} \int dv_i \, v_i f_B(v_i) ( 1 - x_i v_i^2)^{\frac{3}{2}} \ln \bigg| \frac{\zeta + 1}{\zeta - 1} \bigg| \,,
\end{equation}
where we have defined
\begin{equation}
    \zeta \equiv \frac{x}{x_i v_i} \sqrt{v_i^2 - \frac{1}{x_i} + \frac{1}{x}} \,,
\end{equation}
which takes values $\zeta \geq 1$. For each $x_i$, we will first perform the integral over $v_i$. The range of $v_i$ can be found by demanding that $E \leq 0$, corresponding to bound orbits, and that the integrand of Eq.~(\ref{eq:rdot}) be positive. There are two sub-cases, depending on whether $x_i < x$ or $x_i > x$.\\

\noindent \textbf{Case 1) $\mathbf{x_i < x}$\,:} Here, the particle initially moves \textit{outward}, corresponding to the situation where $E_K> E_P$ at turn-around. We have already demonstrated that this regime is not relevant for our stellar-mass PBHs of interest, and so we will not consider it further beyond commenting that this case is what gives rise to the intermediate $\rho \propto r^{-3/2}$ power law scaling found in Ref.~\cite{Eroshenko:2016}. We thus do not expect to see this behavior in our density profile.\\

\noindent \textbf{Case 2) $\mathbf{x_i > x}$\,:} Here, the particle moves always inwards since $E_K < E_P$ at turn-around. The limits of integration for $v_i$ are
\begin{equation}
    v_i^{\rm min} = 0 \,, \,\,\, v_i^{\rm max} = \sqrt{\frac{x}{x_i(x+x_i)}} \,.
\end{equation}
Though in Fig.~1 of the main text we evaluate Eq.~(\ref{eq:rhoeq}) numerically with the distribution of Eq.~(\ref{eq:boltzmann}), for the sake of obtaining an analytic expression here we can approximate $f_B(v_i) d^3 v_i \simeq \delta^{(3)}(\vec{v}_i) d^3v_i$, which is reasonable since we are thoroughly in the regime where potential energy dominates kinetic. In this case, Eq.~(\ref{eq:rhoinprogress}) becomes\footnote{Technically, the upper limit should be $x(t_{\rm eq})$. This is sufficiently large, however, that little error is incurred in approximating $x(t_{\rm eq}) \rightarrow \infty$.}
\begin{equation}
    \rho(x) \simeq \frac{2}{\pi} \int_x^\infty dx_i \frac{x_i}{x^{3/2}} \frac{\rho_i(x_i)}{\sqrt{x_i - x}} \,.
\end{equation}
In the region $x> x_{\rm ta}(t_{\rm kd}) \equiv x_{\rm kd}$ where $\rho_i(x_i) = \rho_{\rm sp}(x_i)$, with the initial mini-spike profile 
\begin{equation}\label{eq:rhospike}
    \rho_{\rm sp}(r) = \frac{\rho_{\rm eq}}{2} f_{\rm DM} \left( \frac{r_{\rm ta}(t_{\rm eq})}{r} \right)^{9/4} \,,
\end{equation}
this evaluates to 
\begin{equation}\label{eq:rhoradfinal}
    \rho(x) \simeq \alpha \, \rho_{\rm kd} \left( \frac{x_{\rm kd}}{x} \right)^{9/4} \,,
\end{equation}
with $\alpha \equiv \sqrt{\frac{4}{\pi}} \frac{\Gamma(3/4)}{\Gamma(5/4)} \simeq 1.526$. Thus, we see that this effect serves to amplify the density everywhere by $\sim 53\%$. In particular, the power law scaling $\rho \propto x^{-9/4}$ is retained. 

%%%%%%%%%%%%%%%%%%%%%%%%%%%%%%%%%%%%%%%%%
%%%%%%%%%%%%%%%%%%%%%%%%%%%%%%%%%%%%%%%%%
\section{Constraints on DM Annihilations}\label{sec:appendixB}

DM self-annihilations in the density spike proceed at a rate \cite{Boucenna:2017}
\begin{equation}
    \Gamma_{\rm ann} = \frac{\expval{\sigma v}}{m_{\rm DM}^2} \int d^3 r \, \rho(r)^2 \,,
\end{equation}
which for the annihilation-modified density profile
\begin{equation}\label{eq:annmodifiedrho}
    \rho_{\rm ann}(x) = \begin{cases} \rho_{\rm max} & 1 < x \leq x_{\rm ann} \,, \\ \rho_{\rm max} \left(\frac{x_{\rm ann}}{x} \right)^{9/4} & x_{\rm ann} \leq x \leq x_{\rm eq} \,, \end{cases} 
\end{equation}
can be explicitly evaluated as
\begin{equation}
    \Gamma_{\rm ann} = \frac{4 \pi \expval{\sigma v} r_{\rm ann}^3 \rho_{\rm max}^2}{m_{\rm DM}^2} \,.
\end{equation}
The $\gamma$-ray flux from these annihilations contributes to both galactic and extragalactic backgrounds, and so can be compared against measurements of the isotropic $\gamma$-ray flux. 

Following Ref.~\cite{Cirelli:2012}, we resolve the total flux into galactic and extragalactic components
\begin{equation}
    \frac{d\Phi_\gamma}{d E_\gamma} = 4 \pi \frac{d^2 \Phi_{\rm gal}}{dE_\gamma d\Omega} \bigg|_{\rm min} + \frac{d\Phi_{\rm ex}}{dE_\gamma} \,.
\end{equation}
Note that although the galactic contribution is not isotropic, its minimum still constitutes an irreducible background contribution to the isotropic flux. Explicitly, these fluxes are \cite{Cirelli:2012,Boucenna:2017}
\begin{equation}
    \frac{d^2 \Phi_{\rm gal}}{dE_\gamma d\Omega} = f_{\rm DM}^2 (1-f_{\rm DM}) \frac{\Gamma_{\rm ann}}{M_{\rm BH}} \frac{1}{4\pi} \int_{\rm l.o.s.} ds \frac{dN_\gamma}{dE_\gamma} \rho_{\rm ann}(s) \,,
\end{equation}
and
\begin{equation}
    \frac{d\Phi_{\rm ex}}{dE_\gamma} = f_{\rm DM}^2 (1-f_{\rm DM}) \frac{\Gamma_{\rm ann}}{M_{\rm BH}} \rho_{\rm DM}^0 \int_0^\infty dz \frac{dN_\gamma}{dE_\gamma} \frac{e^{-\tau_{\rm opt}(z)}}{H(z)} \,,
\end{equation}
where $dN_\gamma/dE_\gamma$ describes the number of photons per unit energy produced in the annihilation channel. The first integral is taken over the line of sight (l.o.s.) $s$ while the second is taken over all redshifts. We also note that the density $\rho_{\rm DM}^0$ appearing in the second expression is the total DM density today, and the optical depth $\tau_{\rm opt}(z)$ appears in the second integral to account for the attenuation of high-energy $\gamma$-rays.

Following Ref.~\cite{Boucenna:2017}, we will utilize the fact that these expressions are identical to those for decaying DM provided we make the substitution
\begin{equation}
    \frac{\Gamma_{\rm dec}}{m_{\rm DM}} \rightarrow f_{\rm DM}^2 (1-f_{\rm DM}) \frac{\Gamma_{\rm ann}}{M_{\rm BH}} \,.
\end{equation}
The idea then is to translate the experimental bounds for decaying DM to constraints on $\expval{\sigma v}$. In particular, we consider the bounds of Ref.~\cite{Ando:2015} derived from data from the Large Area Telescope (LAT) onboard the Fermi satellite \cite{Fermi:2014}. The strongest bounds on DM lifetime for much of the $1\,\text{GeV}-1\,\text{TeV}$ mass range come from the $b\bar{b}$ channel and constrain $\tau_{\rm dec} \gtrsim 10^{28}\,\text{s}$. If we take for example $f_{\rm PBH} \simeq 10^{-4}$, then for a $\mathcal{O}(1\text{-}10)\, M_\odot$ PBH we must demand $\expval{\sigma v} \lesssim 10^{-40}\, \text{cm}^3/\text{s}$. 

Such small cross sections correspond to $\rho_{\rm max}$ values which are much too large to bring us within range of our target DM densities. The smaller $\rho_{\rm max}$ values needed to explain the anomalous orbital decays would result in an overproduction of $\gamma$-rays, and so we do not consider them further. We remark, however, that if there are other astrophysical effects which reduce DM density in the halo, then these annihilation profiles could become relevant. 

%%%%%%%%%%%%%%%%%%%%%%%%%%%%%%%%%%%%%%%%%
%%%%%%%%%%%%%%%%%%%%%%%%%%%%%%%%%%%%%%%%%
\section{BH-LMXB Formation}\label{sec:appendixB2}

Here we present an estimate for the number of PBH-star binaries one could expect to find in the Milky Way, as well as an assessment of the extent to which the DM profile may be disrupted during BH-LMXB formation. 

\subsection{Capture Estimate}

We presume formation occurs predominantly through the dynamical capture channel, with energy dissipation provided by dynamical friction due to the PBH's DM spike. Intuitively, the number of binaries should be given by the number of PBH-star encounters $N_{\rm enc}$ with impact parameter less than same critical $b_{\rm max}$ times the probability that a given encounter results in the formation of a bound system $P_{\rm form}$,
\begin{equation}
    N_{\rm binaries} = N_{\rm enc} P_{\rm form} \,.
\end{equation}

The encounter rate per unit volume is 
\begin{equation}\label{eq:encrate}
    \Gamma_{\rm enc} = n_{\rm PBH} \, n_* \, \sigma \, v_{\rm rel} \,,
\end{equation}
where $n_{\rm PBH}$ is the number density of PBHs, $n_* \simeq 0.1 \, \text{pc}^{-3}$ is the local stellar density, $\sigma$ is the interaction cross section, and $v_{\rm rel} \simeq v_* \sim 100\, \text{km}/\text{s}$ is the relative velocity between the bodies. The PBH number density can be expressed in terms of the fraction of DM constituted by PBHs, $f_{\rm PBH} = \rho_{\rm PBH}/\rho_{\rm DM}$, as
\begin{equation}
    n_{\rm PBH} = \frac{\rho_{\rm PBH}}{M_{\rm BH}} = \frac{f_{\rm PBH} \, \rho_{\rm DM}}{M_{\rm BH}} \,,
\end{equation}
with $\rho_{\rm DM} = 0.3 \, \text{GeV}/\text{cm}^3$ the local DM density. Note that we take $f_{\rm PBH} \lesssim 10^{-3}$, in accordance with the constraints on stellar-mass PBHs \cite{Green:2020}. We estimate the interaction cross section as 
\begin{equation}
    \sigma \simeq \pi b_{\rm max}^2 \,,
\end{equation}
where $b_{\rm max}$ is the maximum impact parameter for which the star is gravitationally influenced by the PBH. This can be estimated by solving for the distance at which the star's kinetic energy $K_* \simeq \frac{1}{2} M_* v_{\rm rel}^2$ is comparable to the gravitational potential energy $U_*$ due to the PBH and surrounding DM spike,
\begin{equation}
    U_*(r) = - \frac{G (M_{\rm BH} + M_{\rm sp}) M_*}{r} \,,
\end{equation}
%\begin{equation}
%    U_*(r) = - \frac{G(M_{\rm BH} + M_{\rm sp}(r)) M_*}{r} \,,
%\end{equation}
%where $M_{\rm sp}(r)$ is the mass enclosed by the DM mini-spike
%\begin{equation}
%    M_{\rm sp}(r) = 4\pi \int_r dr' \, r'^{\,2} \rho(r') \,,
%\end{equation}
%and $\rho(r)$ is given by Eq.~(\ref{eq:rhomatt}). 
where the mass of the PBH-DM spike system $(M_{\rm BH} + M_{\rm sp})$ is given by
\begin{equation}\label{eq:boundmass}
    M_{\rm bound}(t) = M_{\rm bound}(t_{\rm eq}) \left( \frac{t}{t_{\rm eq}} \right)^{2/3} \,,
\end{equation}
evaluated today. Equating $|U_*(b_{\rm max})| = K_*$ and solving for $b_{\rm max}$ gives
\begin{equation}
    b_{\rm max} = \frac{2 G (M_{\rm BH} + M_{\rm sp})}{v_{\rm rel}^2} \,.
\end{equation}
For a rough estimate of the number of encounters, we can multiply the rate per unit volume of Eq.~(\ref{eq:encrate}) by the volume of the Milky Way ($\sim 2 \times 10^{11} \, \text{pc}^3$) and the time since reionization ($\sim 10^{10}\, \text{yr}$). 

Next, we need the probability that a given encounter will actually result in the formation of a binary. In order for the star to be captured, dynamical friction must remove a sufficient amount of kinetic energy such that the system can become gravitationally bound, $E_* < 0$. Recall that by definition, though, $E_* = K_* + U_* = 0$ at $b_{\rm max}$. Thus, for $b \lesssim b_{\rm max}$ essentially any encounter will result in $E_* < 0$, and we can approximate the probability as $1$ provided $b \lesssim b_{\rm max}$. A conservative estimate for the number of binaries is then simply the number of encounters with $b \lesssim b_{\rm max}$,
\begin{equation}
    N_{\rm binaries} \simeq N_{\rm enc}(b \lesssim b_{\rm max}) \,.
\end{equation}
Numerically, this comes out to be
\begin{equation}
    N_{\rm binaries} \sim 10^6 \left( \frac{f_{\rm PBH}}{10^{-5}} \right) \left( \frac{M_{\rm BH}}{5\, M_\odot} \right) \,,
\end{equation}
which can indeed be non-trivial.

Of course there are many ways by which one could improve this rough order-of-magnitude estimate. For example, one should really be integrating the rate of Eq.~(\ref{eq:encrate}) over time since reionization, as well as taking into account the time evolution of $(M_{\rm BH} + M_{\rm sp})$, rather than simply evaluating today. One should also keep in mind that due to dynamical friction, these binaries have finite lifetimes. See Ref.~\cite{Esser:2022} for more ideas on improving the capture estimate. Here we simply wish to show that $N_{\rm binaries}$ can be non-vanishing, and so these rough approximations suffice for the purposes of this paper.

%%%%%%%%%%%%%%%%%%%%%%%%%%%%%%%%%%%%%%%%%
%%%%%%%%%%%%%%%%%%%%%%%%%%%%%%%%%%%%%%%%%
\subsection{Mini-Spike Disruption} 

Since dynamical friction of the star moving through the DM mini-spike is the energy dissipation mechanism responsible for binary formation, a potential concern one may have is that too much of the halo may be stripped away during formation, destroying the mini-spike. This was found to occur in Ref.~\cite{Kavanagh:2018} in the context of extremely eccentric BH-BH binaries when the PBHs had comparable masses. Though we expect this not to be a danger here, since the star is much lighter than the PBH and the final orbits have nearly vanishing eccentricity, it is still worthwhile to verify this is the case.

To estimate the extent to which the DM mini-spike is disrupted during the BH-star binary formation, we can compare the mini-spike's binding energy to the energy lost to dynamical friction \cite{Kavanagh:2020}. To calculate the binding energy of the DM distribution, we consider the work required to assemble the spike by adding spherical shells of DM at successively larger radii. The potential energy of each shell is given by
\begin{equation}\label{eq:dU}
    dU = - \frac{G(M_{\rm BH} + M_{\rm sp}(r))}{r} 4\pi r^2 \rho(r)  dr \,.
\end{equation}
The binding energy of the DM distribution $\Delta U(r)$ is then obtained by integrating the right-hand side.

Next, we want to compute the work done by dynamical friction in binding the stellar body. The force due to dynamical friction is 
\begin{equation}
    F_{\rm df} = 4\pi G^2 \mu^2 \, \rho(r) \, \ln \Lambda / v_{\rm rel}^2 \,,
\end{equation}
where $\mu = M_{\rm BH} M_*/(M_{\rm BH} + M_*)$ is the reduced mass and $\Lambda$ is the Coulomb logarithm, which we approximate as $\Lambda \simeq \sqrt{M_{\rm BH}/M_*}$. The work done along some trajectory $\gamma$ is then given by the line integral
\begin{equation}
    W_{\rm df} = \int_\gamma d\vec{s} \cdot F_{\rm df} \,.
\end{equation}

Suppose the star traverses a chord of the DM spike of length $d$, after which it has lost a sufficient amount of energy to begin orbiting at a radius $R$. The associated work is
\begin{equation}
    W_{\rm df} = \frac{8\pi G^2 \mu^2 \ln \Lambda}{v_{\rm rel}^2} \int_0^{d/2}
 dx \, \rho(\sqrt{x^2 + R^2 - d^2/4}) \,,
\end{equation}
with $\rho(r)$ evaluated at $r= \sqrt{x^2 + R^2 + d^2/4}$. This is of course maximized in the limit $d \rightarrow 2R$. Even in this limit, we find generically that $W_{\rm df}/|\Delta U| \ll 1$, and so minimal energy is lost to dynamical friction as compared with the binding energy of the spike.

%%%%%%%%%%%%%%%%%%%%%%%%%%%%%%%%%%%%%%%%%
%%%%%%%%%%%%%%%%%%%%%%%%%%%%%%%%%%%%%%%%%
\section{Feedback from Dynamical Friction}\label{sec:appendixC}

Here we analyze how dynamical friction and heating due to gravitational scattering with the companion star redistribute DM density in the halo. We follow the strategy of Ref.~\cite{Kavanagh:2020}, who presented an analogous calculation in the context of a BH binary system. The idea is to track how the phase space distribution of the DM evolves as energy is injected by the orbiting compact object. 

Given the parameters in Table~1 of the main text, it is justified to assume the orbital parameters to evolve on a timescale much greater than the orbital period $P$. We also take the DM halo to be spherically symmetric and isotropic, a reasonable assumption given that both LMXBs have nearly circular orbits. These assumptions allows us to characterize the DM by an equilibrium phase space distribution
\begin{equation}
    f(\mathcal{E}) = m_{\rm DM} \frac{dN}{d^3r \, d^3 v} \,,
\end{equation}
which is a function of energy per unit mass alone,
\begin{equation}
    \mathcal{E} = \frac{v^2}{2} + \Phi(r) \,.
\end{equation}
The Newtonian gravitational potential $\Phi(r)$ receives contributions from both the PBH and the DM halo, 
\begin{equation}
    \Phi(r) = - \frac{G M_{\rm BH}}{r} + \Phi_{\rm halo}(r) \,.
\end{equation}
For a given halo density profile $\rho(r)$, the corresponding $\Phi_{\rm halo}$ can be obtained from the Poisson equation $\nabla^2 \Phi_{\rm halo} = 4 \pi G \rho$ by first computing the mass enclosed in a radius $r$
\begin{equation}
    M_{\rm halo}(r) = 4\pi \int_0^r dr' \, {r'}^{\,2} \rho(r') \,, 
\end{equation}
and then computing the potential as
\begin{equation}
    \Phi_{\rm halo}(r) = -G \int_r^\infty dr' \, \frac{M_{\rm halo}(r')}{{r'}^{\,2}} \,.
\end{equation}
The distribution function is then given by Eddington's inversion formula \cite{Binney:2008}
\begin{equation}\label{eq:inversion}
    f(\mathcal{E}) = \frac{1}{\sqrt{8} \pi^2} \int_\mathcal{E}^0 \frac{d\Phi}{\sqrt{\Phi - \mathcal{E}}} \frac{d^2 \rho}{d \Phi^2} \,.
\end{equation}

Though the DM halo in its entirety is more massive than the PBH by the end of matter domination, the mass contained within the orbital radius of the companion star $r_{\rm orb}$ is smaller by several orders of magnitude. Thus, it is appropriate to approximate the gravitational potential by that of the PBH, $\Phi(r) \simeq - G M_{\rm BH}/r$. This is convenient since it eliminates the need to update $\Phi$ as the halo is perturbed. Further, it allows us to evaluate Eq.~(\ref{eq:inversion}) explicitly. Presuming a generic power law profile $\rho = \rho_0 (r_0/r)^\gamma$, we can write 
\begin{equation}
    \rho(\Phi) = \rho_0 \left( \frac{r_0}{G M_{\rm BH}} \right)^\gamma (-\Phi)^\gamma \,,
\end{equation}
in which case
\begin{equation}
    f_i(\mathcal{E}) = \frac{\gamma(\gamma-1)}{(2\pi)^{3/2}} \frac{\Gamma(\gamma-1)}{\Gamma(\gamma-\frac{1}{2})} \rho_0 \left( \frac{r_0}{G M_{\rm BH}} \right)^\gamma |\mathcal{E}|^{\gamma - 3/2} \,.
\end{equation}
This should be regarded as the ``initial'' DM phase space distribution, not in a temporal sense but rather in that it describes the distribution before the gravitational interactions with the companion star are taken into account. Given a distribution function $f(\mathcal{E})$, the density profile can be obtained as
\begin{equation}\label{eq:recoverrho}
    \rho(r) = 4 \pi \int_0^{v_{\rm max}} dv \, v^2 f(\mathcal{E}) \,,
\end{equation}
with $v_{\rm max} = \sqrt{- 2 \Phi(r)}$. The idea then is to evolve $f_i(\mathcal{E}) \rightarrow f(\mathcal{E})$ and then substitute the result into Eq.~(\ref{eq:recoverrho}) in order to find the final density profile in the halo. 

To that end, we introduce the density of states $g(\mathcal{E})$, computed as
\begin{equation}\label{eq:densityofstates}
\begin{split}
    g(\mathcal{E}) & = \int d^3r \, d^3v \, \delta(\mathcal{E} - \mathcal{E}(r,v))\\
    & = \sqrt{2} \pi^3 G^3 M_{\rm BH}^3 \left|\mathcal{E} \right|^{-5/2} \,.
\end{split}
\end{equation}
This allows us to write the number of particles with energy in the interval $(\mathcal{E},\mathcal{E}+d\mathcal{E})$ as
\begin{equation}\label{eq:Neq}
    N(\mathcal{E}) d\mathcal{E} = \frac{1}{m_{\rm DM}} g(\mathcal{E}) f(\mathcal{E}) d\mathcal{E} \,.
\end{equation}
Let $\mathcal{P}_\mathcal{E}(\Delta \mathcal{E})$ be the probability per orbit for a DM particle of energy $\mathcal{E}$ to scatter with the companion star, gaining an energy $\Delta \mathcal{E}$ in the process. Then we can write the change in the number of particles with energy $\mathcal{E}$ over the course of a single orbit as
\begin{equation}
\begin{split}
    \Delta N(\mathcal{E}) = & - N(\mathcal{E}) \int d(\Delta \mathcal{E}) \, \mathcal{P}_\mathcal{E}(\Delta \mathcal{E}) \\
    & + \int d(\Delta \mathcal{E}) \, N(\mathcal{E} - \Delta \mathcal{E}) \mathcal{P}_{\mathcal{E}- \Delta \mathcal{E}}(\Delta \mathcal{E}) \bigg] \,,
\end{split}
\end{equation}
where the first term describes the depletion of particles initially of energy $\mathcal{E}$ that have been upscattered to $\mathcal{E}+ \Delta \mathcal{E}$ and the second term describes replenishment from particles of initial energy $\mathcal{E}-\Delta \mathcal{E}$ upscattered to $\mathcal{E}$. Using Eq.~(\ref{eq:Neq}), the change in the distribution function becomes 
\begin{equation}\label{eq:deltaf}
\begin{split}
    \Delta & f(\mathcal{E}) = - f(\mathcal{E}) \int d(\Delta \mathcal{E}) \, \mathcal{P}_\mathcal{E}(\Delta \mathcal{E}) \\
    & + \int d(\Delta \mathcal{E}) \,\left( \frac{g(\mathcal{E} - \Delta \mathcal{E}) }{g(\mathcal{E})} \right) f(\mathcal{E} - \Delta \mathcal{E}) \mathcal{P}_{\mathcal{E}- \Delta \mathcal{E}}(\Delta \mathcal{E}) \,.
\end{split}
\end{equation}

To proceed further, we need to know the explicit form of $\mathcal{P}_\mathcal{E}(\Delta \mathcal{E})$. This can be derived using results in Appendix L of Ref.~\cite{Binney:2008}. Consider a scattering event between the orbiting companion star and a DM particle with impact parameter $b$, the geometry for which is depicted in Fig.~7 of Ref.~\cite{Kavanagh:2020}. Due to this encounter, the parallel component of the star's velocity $\nu$ changes as
\begin{equation}
    \Delta \nu_{\parallel} \simeq - \frac{m_{\rm DM}}{M_*} \frac{2 \nu}{1 + b^2 / b_{\rm 90}^2} \,,
\end{equation}
where we have approximated the DM's velocity as negligible relative to that of the star and defined $b_{90}$ as the impact parameter required to produce a $90^\circ$ deflection of the DM, $b_{90} \simeq G M_* / \nu^2$. The resultant change in the star's energy is $\Delta E_* \simeq M_* \nu \Delta \nu_\parallel$, which by conservation of energy $\Delta E_* + m_{\rm DM} \Delta \mathcal{E} = 0$ implies that the DM particle acquires an energy of
\begin{equation}
    \Delta \mathcal{E}(b) = \frac{2 \nu^2}{1+b^2/b_{90}^2} \,.
\end{equation}
This enters into the scattering probability as 
\begin{equation}
\begin{split}
    \mathcal{P}_\mathcal{E}(\Delta \mathcal{E}) & = \frac{1}{g(\mathcal{E})} \int d^3r \, d^3v \, \delta(\mathcal{E} - \mathcal{E}(r,v)) \, \delta (\Delta \mathcal{E} - \Delta \mathcal{E}(b)) \\
    & = \frac{4\pi}{g(\mathcal{E})} \int d^3r \, \sqrt{2\mathcal{E} - 2\Phi(r)} \left( \frac{b_{90}^2 \nu^2}{b_* \, \Delta \mathcal{E}^2} \right) \delta (b - b_* ) \,,
\end{split}
\end{equation}
where we have introduced $b_* = b_{90}\sqrt{\frac{2 \nu^2}{\Delta \mathcal{E}}-1}$ as the root of the remaining delta function. From this, we see that the spatial integration should be carried out over a torus with major and minor radii $r_{\rm orb}$ and $b_*$, respectively. After performing the trivial integral over the azimuthal angle, it is convenient to perform a change of variables $(r,\theta) \rightarrow (b,\varphi)$, related as
\begin{subequations}
\begin{equation}
    r = \sqrt{ r_{\rm orb}^2 + b^2 + 2 r_{\rm orb} b \cos \varphi} \,,
\end{equation}
\begin{equation}
    \sin \theta = \frac{r_{\rm orb} + b \cos \varphi}{\sqrt{ r_{\rm orb}^2 + b^2 + 2 r_{\rm orb} b \cos \varphi}} \,.
\end{equation}
\end{subequations}
By computing the Jacobian for the transformation, one can show that $r dr d\theta = b \,db \,d\varphi$. Using the delta function to perform the integral, the result is
\begin{equation}
\begin{split}
    \mathcal{P}_\mathcal{E}(\Delta \mathcal{E}) & \simeq \frac{8\pi^2 b_{90}^2 \nu^2}{g(\mathcal{E})\, \Delta \mathcal{E}^2} \\
    & \times \int d\varphi \, r(b_*, \varphi) \, \sin \theta(b_*,\varphi) \, \sqrt{2\mathcal{E} - 2 \Phi(b_*,\varphi)} \,.
\end{split}
\end{equation}
Since we are in the regime $r_{\rm orb} \gg b_*$, we can expand each factor in the integrand in a power series in $b_*/r_{\rm orb} \ll 1$, keeping only the leading terms. Evaluating the integral and using Eq.~(\ref{eq:densityofstates}) to simplify the answer, we find to leading order
\begin{equation}
    \mathcal{P}_\mathcal{E}(\Delta \mathcal{E}) \simeq \frac{16 r_{\rm orb} b_{\rm 90}^2 \nu^2}{G^3 M_{\rm BH}^3} \frac{|\mathcal{E}|^{5/2}}{\Delta \mathcal{E}^2} \sqrt{\mathcal{E} + \frac{G M_{\rm BH}}{r_{\rm orb}}} \,.
\end{equation}
Using $b_{90} \simeq q r_{\rm orb}$ and $\nu^2 = G M_{\rm BH}(1+q)/r_{\rm orb}$ as well as writing things in terms of the dimensionless radii $x_{\rm orb} = r_{\rm orb}/r_s$ introduced previously, the scattering probability takes the very compact form
\begin{equation}
    \mathcal{P}_\mathcal{E}(\Delta \mathcal{E}) \simeq (8\, q \, x_{\rm orb})^2 (1+q) \frac{|\mathcal{E}|^{5/2}}{\Delta \mathcal{E}^2} \sqrt{\mathcal{E} + \frac{1}{2 x_{\rm orb}}} \,.
\end{equation}

Now equipped with an expression for $\mathcal{P}_\mathcal{E}(\Delta \mathcal{E})$, we return to Eq.~(\ref{eq:deltaf}). Since the system evolves on a timescale much longer than the orbital frequency, we can convert this to a differential equation by writing $\Delta f \simeq P \, \partial f/\partial t$, where $P$ is the orbital period. Solving with the aid of a modified version of the \texttt{HaloFeedback} code of \cite{HaloFeedback:2020}, we plot a sample result in Fig.~1 of the main text. Note that for simplicity, we have considered the companion star orbiting at a fixed radius. For a self-consistent treatment, however, we should really solve the coupled system of differential equations for the evolving orbital radius $\dot{r}_{\rm orb}$ and $\dot{f}$. Since the required numerics are beyond the scope of this work, we maintain self-consistency by restricting evolution to $\lesssim 9000$ orbits, corresponding to a $< 0.5\%$ change in $r_{\rm orb}$. 

\bibliographystyle{utphys3}
\bibliography{references}

\end{document}